\documentclass[preprint]{elsarticle}

\pdfoutput=1
\usepackage{balance,amsmath,amssymb}
\usepackage[utf8]{inputenc} 
\usepackage[T1]{fontenc} 
\usepackage{microtype}
\usepackage{verbatim}
\usepackage{amsmath}
\usepackage{mathtools}

\usepackage{graphicx}

\usepackage[dvipsnames]{xcolor}

\usepackage{dcolumn}
\usepackage{bm}

\usepackage[colorlinks = true,
            linkcolor = blue,
            urlcolor  = blue,
            citecolor =red,
            anchorcolor = blue]{hyperref}
\usepackage{verbatim}
\usepackage{color,ulem}
\usepackage[english]{babel}

\newcommand{\cP}{\mathcal{P}}
\newcommand{\nn}{\nonumber}
\newcommand{\df}{\mathrm{d}}

\usepackage{blindtext}

    \makeatletter
    \def\ps@pprintTitle{%
       \let\@oddhead\@empty
       \let\@evenhead\@empty
       \def\@oddfoot{\reset@font\hfil\thepage\hfil}
       \let\@evenfoot\@oddfoot
    }
    \makeatother

\begin{document}

\title{On the Hydrodynamic Description of Holographic Viscoelastic Models}

\author[1]{Martin Ammon\corref{cor1}}
\ead{martin.ammon@uni-jena.de}

\author[2]{Matteo Baggioli\corref{cor2}}
\ead{matteo.baggioli@uam.es}

\author[1]{Se\'an Gray\corref{cor3}}
\ead{sean.gray@uni-jena.de}

\author[1,3]{Sebastian Grieninger\corref{cor4}}
\ead{sebastian.grieninger@gmail.com}

\author[4]{Akash Jain\corref{cor5}}
\ead{ajain@uvic.ca}

\address[1]{Theoretisch-Physikalisches Institut, Friedrich-Schiller-Universit\"at Jena,
Max-Wien-Platz 1, D-07743 Jena, Germany.}
\address[2]{Instituto de Fisica Teorica UAM/CSIC, c/Nicolas Cabrera 13-15,
Universidad Autonoma de Madrid, Cantoblanco, 28049 Madrid, Spain.}
\address[3]{Department of Physics, University of Washington, Seattle, WA 98195-1560, USA}

\address[4]{Department of Physics \& Astronomy, University of Victoria, PO Box 1700 STN CSC, Victoria, BC, V8W 2Y2, Canada.}

\begin{abstract}
We show that the correct dual hydrodynamic description of homogeneous holographic models with spontaneously broken translations must include the so-called ``strain pressure'' -- a novel transport coefficient proposed recently. Taking this new ingredient into account, we investigate the near-equilibrium dynamics of a large class of holographic models and faithfully reproduce all the hydrodynamic modes present in the quasinormal mode spectrum. Moreover, while strain pressure is characteristic of equilibrium configurations which do not minimise the free energy, we argue and show that it also affects models with no background strain, through its temperature derivatives. In summary, we provide a first complete matching between the holographic models with spontaneously broken translations and their effective hydrodynamic description.
\end{abstract}

\maketitle
\tableofcontents

\section{Introduction}
\label{intro}
Models with broken translational invariance have attracted a great deal of interest in the holographic community in recent years, especially in relation to their hydrodynamic description \cite{PhysRevLett.41.121,PhysRevA.6.2401,Delacretaz:2017zxd,Donos:2019txg,Armas:2019sbe,Donos:2019hpp,Grozdanov:2018ewh,Ammon:2019wci,Ammon:2019apj} and their possible relevance for strange metal phenomenology \cite{Amoretti:2017axe,Amoretti:2018tzw,Davison:2013txa,Delacretaz:2016ivq}. Particular emphasis has been given to the so-called {homogeneous models}, e.g. massive gravity \cite{Vegh:2013sk,Andrade:2013gsa,Baggioli:2014roa,Alberte:2015isw}; Q-lattices \cite{Donos:2013eha,Amoretti:2017frz}; and helical lattices \cite{Andrade:2017cnc,Andrade:2018gqk}, due to their appealing simplicity. 

Despite the sustained activity in the field, there still remain a number of open questions. For instance, it has been unclear what hydrodynamic framework appropriately describes the near-equilibrium dynamics of field theories dual to these models. The authors of \cite{Delacretaz:2017zxd} wrote down a generic theory of linearised hydrodynamics with broken translations (see also \cite{chaikin2000principles,PhysRevA.6.2401}), which has been widely used in holography~\cite{Baggioli:2019abx,Alberte:2017oqx,Ammon:2019wci,Ammon:2019apj,Amoretti:2017frz,Amoretti:2018tzw,Amoretti:2017axe,Amoretti:2019cef,Amoretti:2019kuf}. However, the first indication that something was amiss came from \cite{Ammon:2019apj}, in the form of a disagreement between the holographic results and the hydrodynamic predictions of~\cite{Delacretaz:2017zxd} regarding the {longitudinal diffusion} mode. Similarly, \cite{Donos:2019hpp} found inconsistencies between the hydrodynamic theory of~\cite{Delacretaz:2017zxd} and the quasinormal perturbations of a bulk model with explicitly broken translations. Considering these results, it became clear that the understanding of hydrodynamics was lacking some fundamental details needed in order to capture the holographic results. 

Recently, a new fully non-linear hydrodynamic theory for viscoelasticity was proposed in \cite{Armas:2019sbe}. At the linear level, this formulation differs from previous formulations of viscoelastic hydrodynamics due to the presence of an additional transport coefficient, $\mathcal{P}$, called the \textit{lattice}- or \textit{strain pressure}. Physically, $\mathcal{P}$ is the difference between the thermodynamic and mechanical pressures; intuitively, $\mathcal{P}$ can be understood as an additional contribution to the mechanical pressure as a result of working around a uniformly strained equilibrium state. 
In this sense the strain pressure is analogous to the magnetisation pressure which appears in the presence of an external magnetic field \cite{Hartnoll:2007ih,Caldarelli:2016nni}. $\mathcal{P}$ is non-zero in the holographic models mentioned above and, as we illustrate in this paper, is fundamental in order to match the holographic results to hydrodynamics. 

It is misleading, however, to dismiss this new coefficient purely as an artifact of background strain. $\cP$ certainly vanishes in an unstrained equilibrium state that minimises the free energy (as discussed in \cite{Donos:2013cka}), but as we will illustrate in this paper, its temperature dependence still carries vital physical information and affects various modes through $\cP' = \partial_T\cP$. For instance, in scale invariant theories this leads to a non-zero bulk modulus $B = - T\cP'/2$. Hence, the preceding hydrodynamic frameworks would still fall short in capturing the near-equilibrium behavior of holographic models without background strain.

In this paper, we consider the most general isotropic Lorentz violating massive gravity theories in two spatial dimensions \cite{Alberte:2015isw}. The dual field theories correspond to isotropic, conformal, and generically strained viscoelastic systems with spontaneously broken translations. By carefully studying the quasinormal modes in these systems, we illustrate that they are perfectly described by the hydrodynamic framework of~\cite{Armas:2019sbe}. We also build a new thermodynamically stable holographic model with zero background strain. Using this unstrained model, we show that the effects of $\mathcal{P}'$ are still present when $\cP$ vanishes in equilibrium.

\section{Viscoelastic Hydrodynamics}

Let us briefly review the formulation of viscoelastic hydrodynamics from~\cite{Armas:2019sbe}; we will start with the generic constitutive relations for an isotropic viscoelastic fluid, including strain pressure, and write down the linear modes predicted by the hydrodynamic framework. We further extend the work of~\cite{Armas:2019sbe} by discussing thermodynamically stable configurations with zero strain pressure in equilibrium, but with nonzero temperature derivatives, and draw a comparison with the previously known results of~\cite{Delacretaz:2017zxd}. We work in $d=2$ spatial dimensions for simplicity. 

\subsection{Constitutive Relations}

The fundamental ingredients in the theory are the fluid velocity $u^\mu$, temperature $T$, and translation Goldstone bosons $\Phi^I$. We define $e^I_\mu = \partial_\mu\Phi^I$, which is used to further define $h^{IJ} = e^I_\mu e^{J\mu}$, $e_{I\mu} = h^{-1}_{IJ}e^J_\mu$, $h_{\mu\nu} = h^{-1}_{IJ}e^I_\mu e^J_\nu$, and the strain tensor $u_{\mu\nu} = \frac12 (h^{-1}_{IJ} - \delta_{IJ}/\alpha^2) e^I_\mu e^J_\nu$, for some constant $\alpha$. The constitutive relations of an isotropic neutral viscoelastic system, written in a small strain expansion, are given as~\cite{Armas:2019sbe}
\begin{subequations}
\begin{align}
\label{eq:consti}
    T^{\mu\nu}
    &= \left(\epsilon + p
    + T\mathcal{P}' u^{\lambda}{}_{\!\!\lambda} \right) u^\mu u^\nu
    + \left( p + \mathcal{P} u^{\lambda}{}_{\!\!\lambda} \right) \eta^{\mu\nu}
    + \mathcal{P} h^{\mu\nu} \nn\\
    - \eta\, &\sigma^{\mu\nu}
    - \zeta\, P^{\mu\nu} \partial_\rho u^\rho
    - 2G\, u^{\mu\nu}
    - (B-G)\, u^{\lambda}{}_{\!\!\lambda} h^{\mu\nu}\,,
\end{align}
with the thermodynamic identities $\mathrm{d}p = s\, \mathrm{d}T$, $\epsilon = Ts - p$ and $P^{\mu\nu} = \eta^{\mu\nu} = u^\mu u^\nu$. Here $p$ and $\mathcal{P}$ are the thermodynamic and strain pressures respectively; $\epsilon$ and $s$ are energy and entropy densities; and $G$ and $B$ are the shear and bulk moduli. $\sigma^{\mu\nu} = 2P^{\rho(\mu}P^{\nu)\sigma}\partial_{\rho}u_{\sigma} - P^{\mu\nu} \partial_\rho u^\rho$ is the fluid shear tensor, while $\eta$ and $\zeta$ are shear and bulk viscosities. All the coefficients appearing here are functions of $T$; prime denotes derivative with respect to $T$ for fixed $\alpha$. Dynamical evolution of $u^\mu$ and $T$ is governed by the energy-momentum conservation equation $\partial_\mu T^{\mu\nu} = 0$; these are accompanied by the configuration (Josephson) equations for the Goldstones
\begin{equation}
\label{eq:config}
    u^\mu e^I_\mu
    = \frac{h^{IJ}}{\sigma}
    \partial_\mu \left( \mathcal{P} e^{\mu}_J
    \,{-}\, (B{-}G) u^{\lambda}{}_{\!\!\lambda} e^\mu_J 
    \,{-}\, 2G u^{\mu\nu} e_{J\nu}
    \right),
\end{equation}
where $\sigma$ is a dissipative coefficient characteristic of spontaneously broken translations.
\label{eq:hydro_eqns}
\end{subequations}

\subsection{Linear Modes}

The $\mathcal{P}$ dependent terms in \eqref{eq:hydro_eqns} have important consequences for the low energy dispersion relation of the hydrodynamic modes. In summary, around an equilibrium state with $u^\mu = \delta^\mu_t$, $T=T_0$, and $\Phi^I = \alpha\, x^I$, we find two pairs of sound modes, one each in longitudinal and transverse sectors, and a diffusion mode in the longitudinal sector 
\begin{equation}\label{disp}
    \omega = \pm v_{\parallel,\perp} k 
    - \frac{i}{2}\Gamma_{\parallel,\perp} k^2
    + \ldots\,,\quad 
    \omega = -iD_\parallel k^2 + \ldots\,.
\end{equation} 
The sound velocities $v_{\parallel,\perp}$, attenuation constants $\Gamma_{\parallel,\perp}$, and diffusion constant $D_\parallel$ are given as
\begin{gather}
    v_\perp^2
    = \frac{G}{\chi_{\pi\pi}}\,,\qquad 
    v_\parallel^2 
    = \frac{(s+\mathcal{P}')^2}{s'\chi_{\pi\pi}}
    + \frac{B+G-\mathcal{P}}{\chi_{\pi\pi}}\,,
    \quad 
  \Gamma_\perp
  = \frac{\eta}{\chi_{\pi\pi}}
  +\frac{G}{\sigma}\frac{s^2 T^2}{\chi_{\pi\pi}^2}\,, \nonumber\\
  D_\parallel
  = \frac{s^2}{\sigma s'}
  \frac{B+G-\cP}{\chi_{\pi\pi}v_\parallel^2}\,,
 \quad
  \Gamma_\parallel
  = \frac{\eta+\zeta}{\chi_{\pi\pi}}
  +\frac{T^2s^2v_\parallel^2}{\sigma\chi_{\pi\pi}}
  \left(1 - \frac{s+\cP'}{Ts'v_\parallel^2}\right)^2 \,.
  \label{eq:modes}
\end{gather}
Here $\chi_{\pi\pi}= \epsilon + p +\mathcal{P}$ is the momentum susceptibility;\footnote{The observation that $\chi_{\pi\pi}\neq \epsilon + p$ in generic holographic models of viscoelasticity (i.e. that the thermodynamic and mechanical pressures are not necessarily equal) was first made in~\cite{Alberte:2017oqx}.} all functions are evaluated at $T = T_0$. Note that the pair of transverse sound modes are not present when $G=0$; instead, they are replaced by a single shear diffusion mode $\omega = -i D_\perp k^2$ with $D_\perp = \eta/\chi_{\pi\pi}$.\footnote{The limit $G \rightarrow 0$ is subtle and must be performed at the level of the transverse sector dispersion relations, $
    \omega^2 sT +  \left( 1 + \frac{\omega Ts}{i\sigma}  \right)
    \left( \omega^2\cP - k^2 G \right)
    + i \omega k^2  \eta =0$.}
We can obtain formulas for various coefficients appearing in \eqref{eq:hydro_eqns} in terms of the free-energy density $\Omega$, stress-tensor one-point function, and (up to contact-terms) retarded two point functions
\begin{gather}
    \epsilon = \langle T^{tt} \rangle\,, \quad
    p = -\Omega\,, \quad
    \mathcal{P} = \langle T^{xx} \rangle + \Omega\,,\quad
    \chi_{\pi\pi}v_\parallel^2 = \lim_{\omega\to0}\lim_{k\to0}
    \mathrm{Re}\,G^R_{T^{xx}T^{xx}}\,,
    \nonumber\\
    G = \chi_{\pi\pi}v_\perp^2 = \lim_{\omega\to0}\lim_{k\to0}
    \mathrm{Re}\,G^R_{T^{xy}T^{xy}}\,,
    \quad 
    \eta = -\lim_{\omega\to0}\lim_{k\to0}
    \frac{1}{\omega}\mathrm{Im}\,G^R_{T^{xy}T^{xy}}\,,
    \nonumber\\
    \frac{(\epsilon+p)^2}{\sigma\chi_{\pi\pi}^2} 
    = \lim_{\omega\to0}\lim_{k\to0}
    \omega\,\mathrm{Im}\,G^R_{\Phi^{x}\Phi^{x}}\,.
    \label{kubos}
\end{gather}
The bulk modulus $B$ can be obtained indirectly using the $v_\parallel^2$ Kubo formula. In the equations in the first line above, the relation between the strain pressure $\mathcal{P}$, thermodynamic pressure $p$, and the mechanical pressure $\langle T^{xx} \rangle$, is manifest.

For our application to holography we shall, in the following, be interested in scale-invariant viscoelastic fluids, wherein $T^\mu{}_{\!\mu} = 0$. This leads to a set of identities
\begin{equation}
    \epsilon = 2(p+\cP)\,,\quad 
    T\cP' = 3\,\cP-2\,B\,, \quad
    \zeta = 0\,.
    \label{conformal-identities}
\end{equation}
Taking derivative of the first relation, we also find the specific heat $c_v = T s' = 2(s+\cP')$. Using the above equations, we can derive a relation between sound velocities, i.e. $v_\parallel^2 = 1/2 + v_\perp^2$~\cite{Esposito:2017qpj}. For scale-invariant theories $v_\perp$ and $\Gamma_\perp$ stay the same as in \eqref{eq:modes}, however the expressions for the longitudinal sector simplify to
\begin{gather}
  \Gamma_\parallel
  = \frac{\eta}{\chi_{\pi\pi}}
  +\frac{T^2s^2 G^2}{\sigma\chi_{\pi\pi}^3v_\parallel^2}\,, \quad
  D_\parallel
  = \frac{Ts^2/\sigma}{s+\cP'}
  \frac{B+G-\cP}{\chi_{\pi\pi} + 2G}\,.
  \label{eq:modes_conformal}
\end{gather}
Interestingly, apart from the implicit dependence in $\chi_{\pi\pi}$, in a scale-invariant viscoelastic fluid only $D_\parallel$ depends explicitly on $\cP$ and $\cP'$, which explains the discrepancy reported in the diffusion mode in~\cite{Ammon:2019apj}. Note that using \eqref{conformal-identities}, the bulk modulus can be rewritten as $B = (3\cP-T\cP')/2$. Consequently, a scale-invariant viscoelastic system only responds to bulk stress if $\cP,\cP'\neq 0$.\footnote{Nevertheless, the compressibility $\beta\equiv (- 1/V)\,\partial T_{xx}/\partial V$ is finite even in the absence of the strain pressure, and in the scale-invariant case it is given by $\beta^{-1}=(3/4) \epsilon$ \cite{Baggioli:2019elg}. It is possible to show that in terms of the compressibility the longitudinal speed can be written as $v_{\parallel}^2=(\beta^{-1}+G)/\chi_{\pi\pi}$ \cite{Ammon:2019apj,Baggioli:2019abx}.}


\subsection{Unstrained Equilibrium Configurations}

Let us now extend the analysis of~\cite{Armas:2019sbe} by considering equilibrium states without background strain, i.e. states where the equilibrium strain pressure is zero, $\cP(T_0)=0$. In such a setup the temperature derivative of the strain pressure need not vanish, hence $\cP'(T_0)\neq0$.\footnote{We will return to this point in further detail below.} Nevertheless, the momentum susceptibility reduces to a familiar expression $\chi_{\pi\pi}= \epsilon+p$. For generic scale-non-invariant theories, we arrive at the modes
\begin{gather}
    v_\perp^2
    = \frac{G}{Ts}\,,\qquad 
    v_\parallel^2 = \frac{(s+\mathcal{P}')^2}{Ts s'} + \frac{B+G}{Ts}\,,
   \quad 
  \Gamma_\perp
  = \frac{\eta}{Ts} + \frac{G}{\sigma}\,,   \nn\\
  D_\parallel = \frac{s}{\sigma T s'} \frac{B+G}{v_\parallel^2}\,,
\quad 
  \Gamma_\parallel = \frac{\eta+\zeta}{Ts}
  + \frac{Tsv_\parallel^2}{\sigma}
  \left(1 - \frac{s+\cP'}{Ts'v_\parallel^2}\right)^2 \,.
  \label{eq:zeroP-modes}
\end{gather}
In the scale-invariant limit, the longitudinal modes further simplify to $v_\parallel^2 = 1/2 + v_\perp^2$ along with
\begin{equation}
    \Gamma_\parallel
    = \frac{\eta}{Ts}
    +\frac{2G^2/\sigma}{Ts + 2G}\,, \quad
    D_\parallel
    = \frac{Ts^2/\sigma}{s+\mathcal{P}'}
    \frac{B+G}{Ts + 2G}\,.
    \label{modes_zeroP}
\end{equation}
The appearance of $\mathcal{P}'$ in the denominator of $D_\parallel$ suggests that the temperature dependence of strain pressure still plays an important role in an unstrained equilibrium configuration. Indeed, $\mathcal{P}'$ is crucial for thermodynamically stable holographic models, as we illustrate below. In the absence of scale invariance, the effects of $\mathcal{P}'$ will also contaminate the expression for the longitudinal sound mode. Other signatures of strain pressure in a scale-invariant viscoelastic system include non-canonical specific heat, $c_v = 2(s+\cP')\neq 2s$, and nonzero bulk modulus $B = - T\cP'/2 \neq 0$.\footnote{Note that~\cite{Armas:2019sbe} assumes $\cP'$ to also vanish in theories with zero strain pressure, leading to zero bulk modulus in scale invariant unstrained theories.}

Comparing our results to~\cite{Delacretaz:2017zxd}, we find that \eqref{eq:zeroP-modes} matches the expressions derived using the hydrodynamic framework of~\cite{Delacretaz:2017zxd} for neutral relativistic viscoelastic fluids only if we further set $\cP' = 0$. As a consequence, the results of~\cite{Delacretaz:2017zxd} do not apply to general unstrained viscoelastic systems with nonzero $\cP'$. Notably, the analysis of~\cite{Delacretaz:2017zxd} can be extended to include certain couplings in the free-energy density that have been switched off therein (see (A.7) of~\cite{Delacretaz:2017zxd}). We find that such couplings are indeed important and precisely capture the effects of nonzero $\cP'$ via the mapping $b = -\cP'/s'$.


\section{Holographic Framework}

\subsection{Holographic Massive Gravity}

We will consider a simple holographic model with $(d+2)$-dimensional Einstein-AdS gravity coupled to $d$ copies of St\"uckelberg scalars $\phi^I$
\begin{equation}\label{action}
    S_{\text{bulk}} 
    \,{=}\! \int \mathrm{d}^{d+2}x \sqrt{-g}
    \left(\frac{R}2
    \,{+}\, \frac{d(d{+}1)}{2\ell^2}
    \,{-}\, m^2 V(\mathcal{I}^{IJ})\right),
\end{equation}
where $\mathcal I^{IJ} = g^{ab}\partial_a\phi^I\partial_b\phi^J$ is the kinetic matrix; $\ell$ is the AdS-radius, which we set to one in the following; and $m$ is a parameter related to the graviton mass. We have set $8\pi G_N/c^4 = 1$. For the isotropic case in $d=2$, we can generically take $V(\mathcal{I}^{IJ}) = V(X,Z)$ where $X = \frac12 \mathrm{tr}\,\mathcal{I}$ and $Z=\text{det}\,\mathcal{I}$ \cite{Baggioli:2014roa,Alberte:2015isw,Baggioli:2019abx}. The scalars $\phi^I$ are \textit{dual} to the boundary operators $\Phi^I$ and break the translational invariance of the dual field theory (see \cite{Nicolis:2015sra} and \cite{Alberte:2015isw} for the specifics of the symmetry breaking pattern). Depending on the boundary conditions imposed on $\phi^I$, this breaking can either be explicit, spontaneous, or pseudo-spontaneous \cite{Andrade:2013gsa,Alberte:2017cch,Ammon:2019wci,Baggioli:2019abx,Armas:2019sbe}. Presently, we shall be interested in models with spontaneously broken translations leading to phonon dynamics in the dual field theory~\cite{Alberte:2017oqx, Ammon:2019apj, Baggioli:2019abx,Baggioli:2019aqf, Baggioli:2019elg}.

We consider a black brane solution of \eqref{action} in Eddington-Finkelstein (EF) coordinates with the metric
\begin{gather}
\mathrm{d}s^2 = \frac{1}{u^2} \left(-f(u)\,\mathrm{d}t^2-2\,\mathrm{d}t\,\mathrm{d}u + \mathrm{d}x^2+\mathrm{d}y^2\right)\,,\,
\label{backg}
\end{gather}
and a radially constant profile for the scalars, $\phi^I = \alpha\, x^I$, for some constant $\alpha$. The radial coordinate $u\in [0,u_h]$ spans from the boundary $u=0$ to the horizon $u=u_h$. The emblackening factor $f(u)$ takes a simple form
\begin{equation}\label{backf}
f(u)= 1 - \frac{u^3}{u_h^3}  
- u^3 \int_{u}^{u_h} \,
\frac{m^2}{\aleph^4}\,V(\alpha^2\,\aleph^2,\alpha^4\,\aleph^4)\,\mathrm{d}\aleph \, .
\end{equation}
Linear perturbations around the black brane geometry capture near-equilibrium finite temperature fluctuations in the boundary field theory~\cite{Alberte:2016xja,Andrade:2019zey,Baggioli:2019abx,Baggioli:2019elg,Baggioli:2019mck}. 

Temperature and entropy density in the boundary field theory are identified with the Hawking temperature and area of the black brane, respectively
\begin{equation}
    T = -\frac{f'(u_h)}{4\pi}
    = \frac{3 -  m^2\, V_h}{4 \pi\, u_h}\,, \qquad 
    s = \frac{2\pi}{u_h^2}\,,
    \label{eq:temp}
\end{equation}
with $V_h = V(u_h^2\alpha^2,u_h^4\alpha^4)$. The free energy density is defined as the renormalised euclidean on-shell action \cite{Skenderis:2002wp}. The expectation value $\langle T^{\mu\nu}\rangle$ can be read off using the leading fall-off of the metric at the boundary. Using the first row of \eqref{kubos}, this leads to the thermodynamic quantities
\begin{gather}
    p 
    = \frac{1}{2u_h^3}
    - \frac{m^2}{u_h^3}\left(\frac{1}{2}V_h - U_h \right)\,, \qquad
    \epsilon
    =\frac{1}{u_h^3} - \frac{m^2}{u_h^3} U_h\,, \quad
    \mathcal{P}
    = \frac{m^2}{u_h^3}
    \left(\frac{1}{2}V_h - \frac32 U_h \right)\,.
    \label{thermo_hol}
\end{gather}
We have defined $U_h = - u_h^3 \int_0^{u_h} \aleph^{-4} V(\alpha^2\aleph^2,\alpha^4\aleph^4) \mathrm{d}\aleph$, assuming $V(X,Z)$ to fall off faster than $\sim\! u^3$ at the boundary.\footnote{\label{foot:alternate-Uh}For potentials that fall of slower than $\sim\!u^3$ near the boundary, such as $V(X)=X^N$ with $N<3/2$, this integral is divergent. Nevertheless, performing holographic renormalisation carefully (see \ref{ap1}), the thermodynamic quantities above can be computed explicitly and amounts to defining $U_h = u_h^3 \int^\infty_{u_h} \aleph^{-4} V(\alpha^2\aleph^2,\alpha^4\aleph^4) \mathrm{d}\aleph$ instead.} Details of holographic renormalisation for these models have been given in \ref{ap1}.
Using the expressions in \eqref{thermo_hol} together with \eqref{conformal-identities}, 
we can find the bulk modulus
\begin{equation}
    B
    = \frac{m^2}{4u_h^3} \left( 3V_h \,{-}\, 9 U_h
    + \frac{u_h\partial_{u_h}\!V_h (m^2 V_h-3)}{m^2
   \left(V_h- u_h\partial_{u_h}\!V_h\right)-3}\right),
    \label{rt}
\end{equation}
Finally, using the results of \cite{Amoretti:2018tzw, Ammon:2019wci, Donos:2019txg}, we can derive a horizon formula for $\sigma$, which reads
\begin{equation}
    \label{eq:sigma_horizon}
    \sigma = \frac{m^2}{2 \alpha^2 u_h^3} \frac{\partial V_h}{\partial u_h},
\end{equation}
and agrees well with the numerical results obtained with the Kubo formula in \eqref{kubos}. The remaining coefficients, $G$ and $\eta$, must be obtained numerically.
 
The non-trivial expression for $\cP$ in \eqref{thermo_hol} indicates the presence of background strain in these holographic models. This is associated with the equilibrium state $\phi^I = \alpha\, x^I$ not being a minimum of free energy~\cite{Donos:2017ihe,Amoretti:2017frz,Alberte:2017oqx}. To wit, using \eqref{thermo_hol} one can check that $\mathrm{d}\Omega/\mathrm{d}\alpha|_T = - \mathrm{d}p/\mathrm{d}\alpha|_T =0$ leads to $\cP = 0$. However, as is evident from \eqref{eq:modes}, the presence of $\cP$ by itself does not lead to any linear instability or superluminality~\cite{Alberte:2017oqx, Ammon:2019apj, Baggioli:2019abx}. Setting $\cP = 0$ in \eqref{thermo_hol}, we can find a thermodynamically favored state $\alpha =\alpha_0$ as a non-zero solution of $V_h = 3U_h$. Notice that $\cP'|_{\alpha = \alpha_0}\neq 0$, which means that strain pressure still plays a crucial role in the dual hydrodynamics through its temperature derivatives, as discussed around \eqref{modes_zeroP}. In particular, these models can have non-zero bulk modulus despite being scale invariant. 

Simple monomial models considered previously in the literature~\cite{Baggioli:2019abx,Alberte:2015isw,Alberte:2017oqx,Andrade:2019zey,Baggioli:2019elg,Baggioli:2019mck}, such as $V(X,Z)=X^N,Z^M$, do not admit $\cP = 0$ states with non-zero $\alpha$.\footnote{However, the would-be preferred state $\alpha=0$ is not a good vacuum of the theory, since the model is strongly coupled around that background \cite{Alberte:2017oqx}. Therefore, in these theories, it is incorrect to compare free energies of states with $\alpha \neq 0$ against the state $\alpha=0$.} The simplest models admitting states with $\cP=0$ have polynomial potentials such as $V(X,Z) = X+\lambda X^2$. Unfortunately, this naive model is plagued by linear instabilities. Nevertheless, it can be used as a toy model to illustrate the importance of $\mathcal{P}'\neq 0$; we return to the details of this model below.


\subsection{Strained Holographic Models} 

Let us first specialize to the strained models with $V(X,Z) = X^N,Z^M$ and $N>5/2$, $M>5/4$ to numerically obtain $G$ and $\eta$, and test the agreement between quasinormal modes and the hydrodynamic predictions. We can compute the full spectrum of quasinormal modes, in both the transverse and longitudinal sectors, using pseudo-spectral methods following \cite{Ammon:2019apj,Baggioli:2019abx,Alberte:2017oqx,Ammon:2016fru,Grieninger:2017jxz}. As we discussed around \eqref{eq:modes_conformal}, the strain pressure does not appear explicitly in the transverse sound modes, leading to the same predictions by \cite{Delacretaz:2017zxd} and \cite{Armas:2019sbe}, modulo the definition of $\chi_{\pi\pi}$. Since the discrepancy in $\chi_{\pi\pi}$ has already been identified and tested against holographic results~\cite{Alberte:2017oqx, Baggioli:2019abx}, here we only focus on the longitudinal sector.

\begin{figure}[h]
    \centering
    \includegraphics[width=0.45\linewidth]{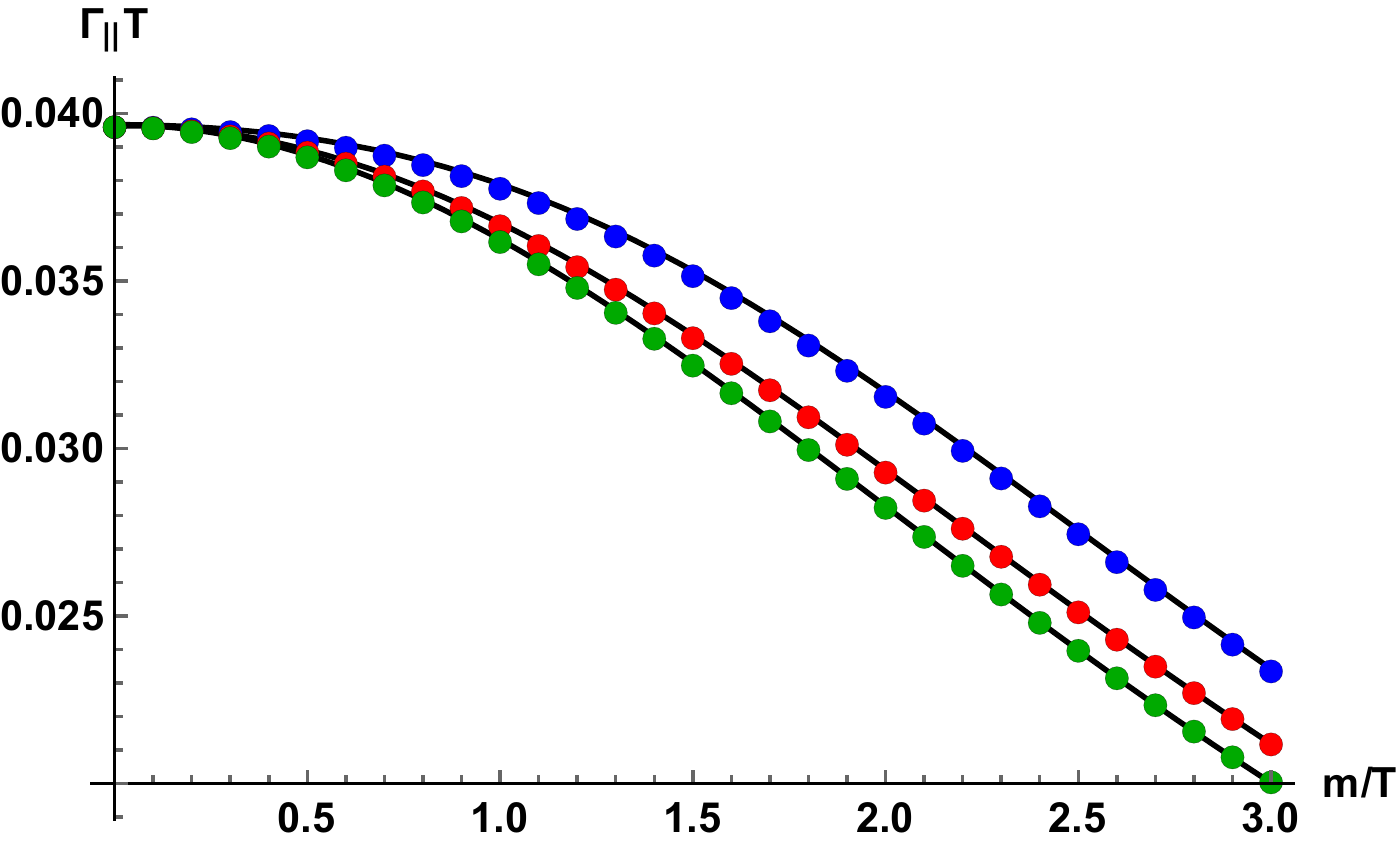} \quad 
    \includegraphics[width=0.45\linewidth]{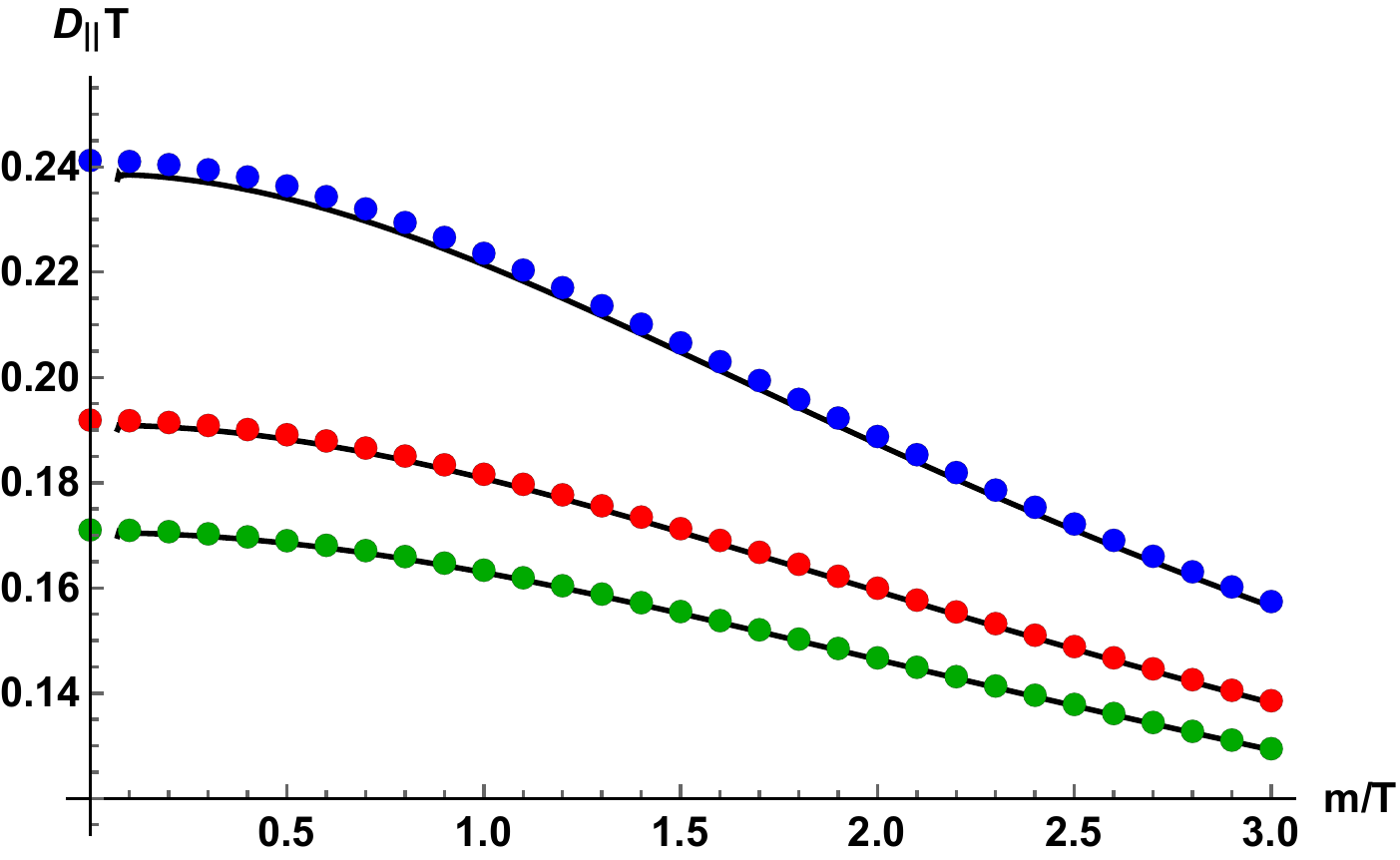}
    
    \caption{$\Gamma_\parallel$ and $D_\parallel$ for $V(X,Z) = X^N$ models for $N=3,4,5$ (from top to bottom) as functions of the dimensionless parameter $m/T$, alongside their hydrodynamic predictions from \eqref{eq:modes} (solid lines).}
    \label{fig_X}
\end{figure}

We start with $V(X,Z)=X^N$ models. Note that $V_h = \alpha^{2N}u_h^{2N}$ and $U_h = \alpha^{2N}u_h^{2N}/(3-2N)$. Using \eqref{eq:temp}-\eqref{eq:sigma_horizon}, we can explicitly find
\begin{gather}
    T = \frac{3 -  m^2 V_h}{4 \pi\, u_h}\,, \qquad 
    s = \frac{2\pi}{u_h^2}\,. \quad
    p = \frac{1}{2u_h^3} \left(
    1 - \frac{2N-1}{2N-3}m^2V_h\right), \nn\\
    \epsilon = \frac{1}{u_h^3} \left(
    1+ \frac{m^2V_h}{2N-3}\right), \quad
    \mathcal{P} = \frac{N}{2N-3}\frac{m^2 V_h}{u_h^3}, \qquad
    \mathcal{P}' = - \frac{4\pi}{u_h^2} \frac{Nm^2 V_h}{3+(2N-1)m^2V_h}, \nn\\
    B = \frac{N m^2 V_h}{2u_h^3} 
    \left(\frac{3}{2N-3} + \frac{3 - m^2 V_h}{3 + (2N-1) m^2 V_h}\right), \quad
    \sigma = \frac{N m^2 V_h}{\alpha^2 u_h^4}, \nn\\
    c_v = \frac{4\pi}{u_h^2} \frac{3 - m^2 V_h}{3 + (2N-1) m^2 V_h}.
    \label{eq:thermo-XN}
\end{gather}
Computing $G$ and $\eta$ numerically using \eqref{kubos}, we can compare the hydrodynamic prediction for the longitudinal attenuation constant $\Gamma_\parallel$ and diffusion constant $D_\parallel$ in \eqref{eq:modes} with the numerical results obtained for the quasinormal modes in the holographic model. The results are shown in fig. \ref{fig_X}. The agreement is extremely good and is valid independent of $N$. We no longer see a discrepancy in the diffusion mode.

Let us now consider models $V(X,Z)=Z^M$. In this case, $V_h = \alpha^{4M}u_h^{4M}$ and $U_h = \alpha^{4M}u_h^{4M}/(3-4M)$. The expressions for thermodynamic quantities remain the same as in \eqref{eq:thermo-XN} but with $N\to2M$. 

\begin{figure}[h]
    \centering
    \includegraphics[width=0.45\linewidth]{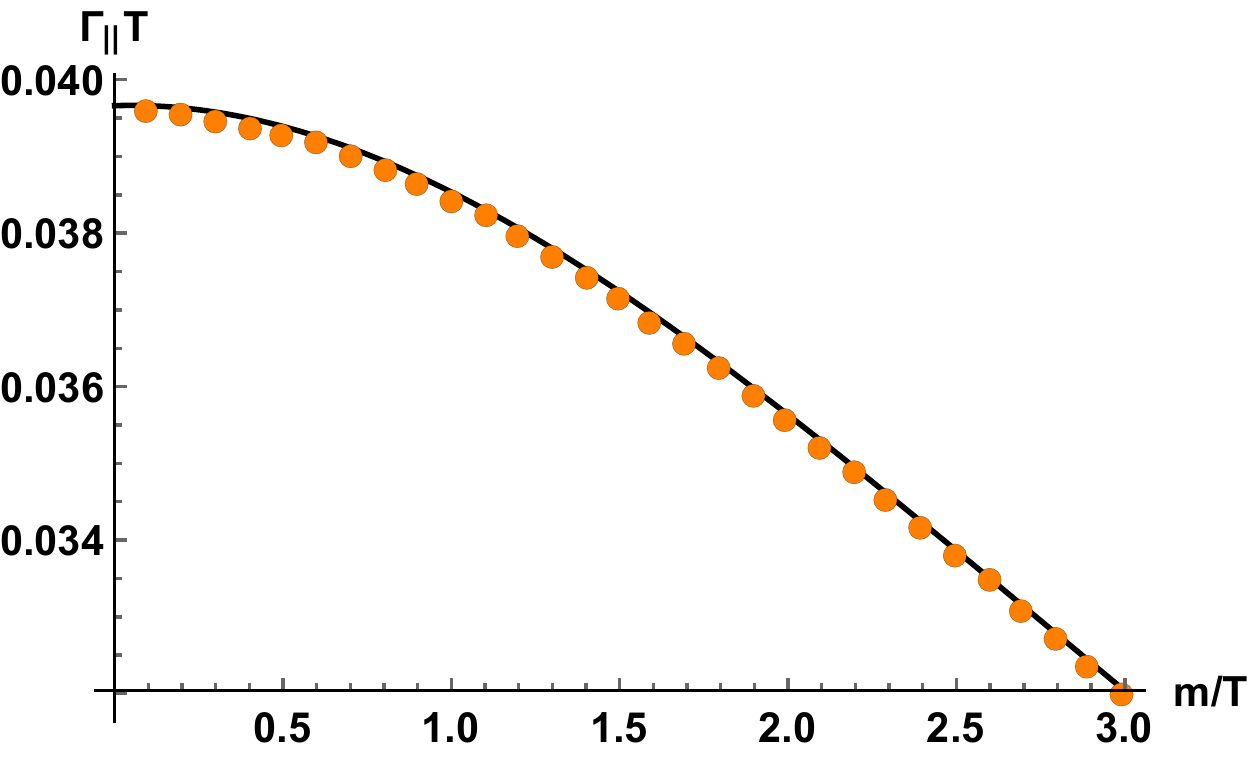}
    \quad \includegraphics[width=0.45\linewidth]{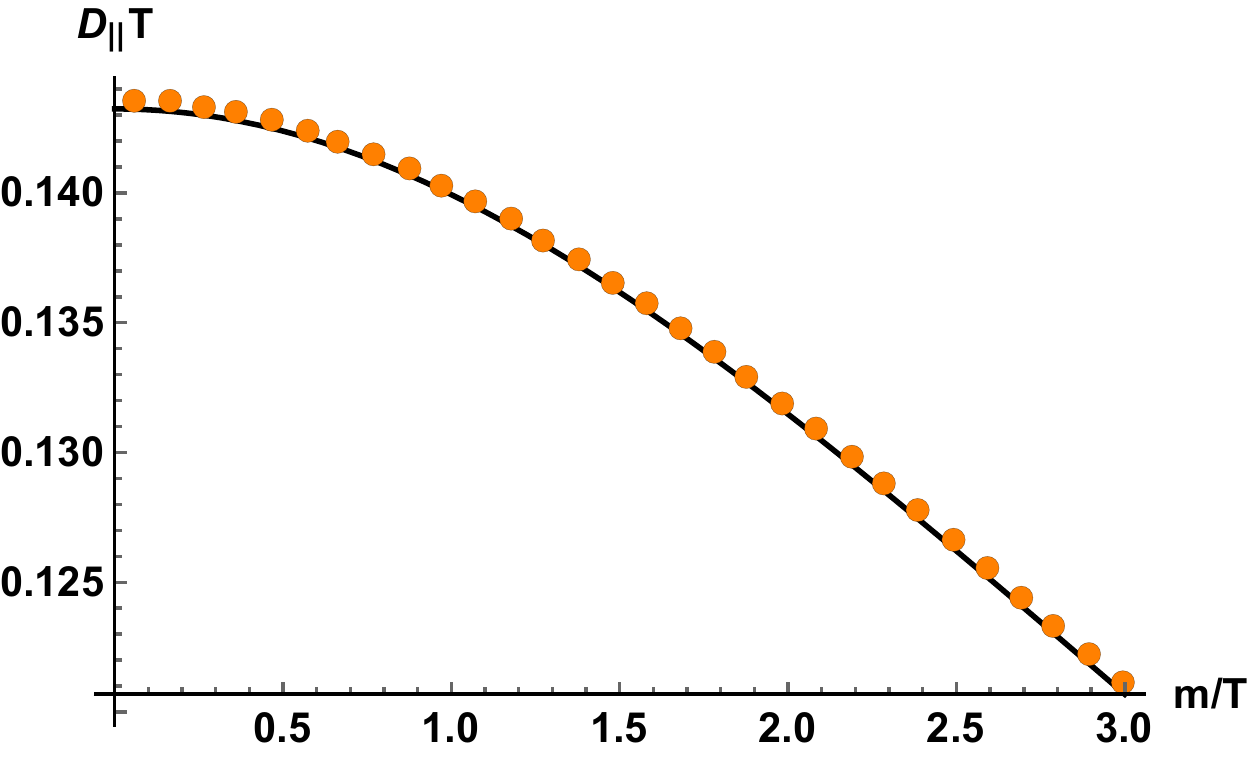}
    \caption{$\Gamma_\parallel$ and $D_\|$ for the model $V(X,Z) = Z^2$, as a function of the dimensionless parameter $m/T$, and the hydrodynamic prediction from \eqref{eq:modes}.}
    \label{fig_Z}
\end{figure}

Generically, $X$-independent potentials $V(X,Z)=V(Z)$ enjoy a larger symmetry group -- the dual field theory is invariant under volume preserving diffeomorphisms, modeling a fluid. These models have $G=0$, leading to the absence of transverse phonons~\cite{Alberte:2015isw}, and $\eta$ saturating the Kovtun-Son-Starinets bound~\cite{Alberte:2016xja}. In fig. \ref{fig_Z} we show a comparison between the hydrodynamic prediction and numerical results for quasinormal modes for $V(X,Z)=Z^2$. The excellent agreement confirms that the hydrodynamic framework of \cite{Armas:2019sbe} is valid for a general class of viscoelastic models with non-zero strain pressure.

\subsection{Unstrained Holographic Models}

In this section, we consider holographic models with zero strain pressure in equilibrium. These are thermodynamically favourable models which admit translationally broken phases that minimise free energy. We will illustrate that even for such models, the strain pressure plays a crucial role in the dual hydrodynamics through its temperature derivatives and hence the hydrodynamic modes are governed by the expressions in eq. \eqref{modes_zeroP}.

Let us consider the simplest model $V(X,Z) = X + \lambda X^2$. As mentioned above, this model is unstable: (I) the shear modulus is negative, (II) the speed of transverse sound is imaginary, and (III) the longitudinal diffusion constant becomes negative at large $m/T$. It can be verified that all the models $V(X,Z)=X^{N_1}+\lambda X^{N_2}$ with spontaneous breaking of translations and $\mathcal{P}=0$ suffer from such linear instabilities, or have ghostly excitations in the bulk.\footnote{More precisely, for models with $N_1<3/2$ the shear modulus is negative; see appendix of \cite{Alberte:2017oqx} for formulae. Hence, also the model considered in \cite{Armas:2019sbe} is dynamically unstable.} Clearly, the model $V(X,Z) = X + \lambda X^2$ cannot describe a stable physical system, but it can be used as a toy example to illustrate the importance of strain pressure. We find that $V_h = \alpha^2u_h^2 + \lambda \alpha^4 u_h^4$ and $U_h = \alpha^2u_h^2 - \lambda \alpha^4 u_h^4$. Setting $\cP$ in \eqref{thermo_hol} to zero, we find the preferred value of $\alpha\neq 0$ to be
\begin{equation}
    \alpha^2 = \frac{1}{2\lambda u_h^2}\,,
\end{equation}
which matches the result of \cite{Amoretti:2017frz} in the zero charge density limit $\rho=0$.\footnote{The notational relationships are $\alpha \equiv k$ and $\lambda\equiv\lambda_2$, where the right-hand sides of the identifications are the notation of \cite{Amoretti:2017frz}. Notice also that eq. (45) in \cite{Amoretti:2017frz} contains typos; it should read
$k^2 I_{Y_1}(0)+2\,\lambda_2 k^4 I_{Y_2}(0)\,-\,\lambda_1\,\rho^2\,k^2\,I_{Z_2}(0)\,=\,0$.} 

We obtain the hydrodynamic parameters
\begin{gather}
    T = \frac{3}{4\pi u_h}\left( 1 - \frac{m^2}{4\lambda} \right), \qquad
    s = \frac{2\pi}{u_h^2} ,\quad 
    p 
    = \frac{1}{2u_h^3} \left( 1 - \frac{m^2}{4\lambda} \right), \nn\\
    \epsilon
    =\frac{1}{u_h^3} \left( 1 - \frac{m^2}{4\lambda} \right), \quad
    \cP' = \frac{-4\pi}{3u_h^2}
    \frac{m^2}{\lambda + 5m^2/12}\,, \qquad
    B = \frac{m^2}{2\lambda u_h^3} 
    \frac{\lambda - m^2/4}{\lambda + 5m^2/12}\,,\nn\\
    \sigma = \frac{2m^2}{u_h^2}\,, \qquad
    c_v = \frac{4\pi}{u_h^2} 
    \frac{\lambda - m^2/4}{\lambda + 5m^2/12}\,.
\end{gather}
Notice that the potential behaves as $\sim\! u^2$ near the boundary, so the alternate definition of $U_h$ given in footnote \ref{foot:alternate-Uh} has to be used in formulas \eqref{thermo_hol}-\eqref{rt}. $G$ and $\eta$ have to be found numerically using \eqref{kubos}. We see that $\cP'\neq 0$ leading to $B\neq 0$ and $c_v \neq 2s$ in these models, as discussed above. 

\begin{figure}[h]
   \centering
    \includegraphics[width=0.47\linewidth]{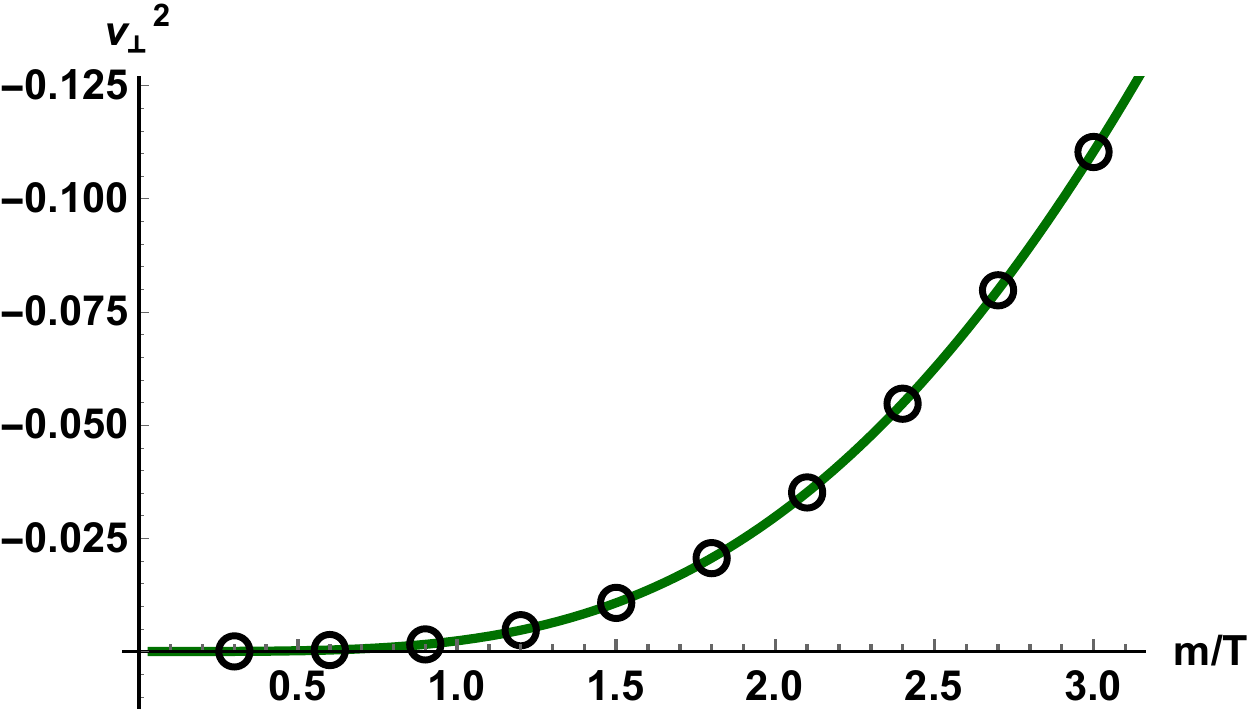}\quad 
     \includegraphics[width=0.43\linewidth]{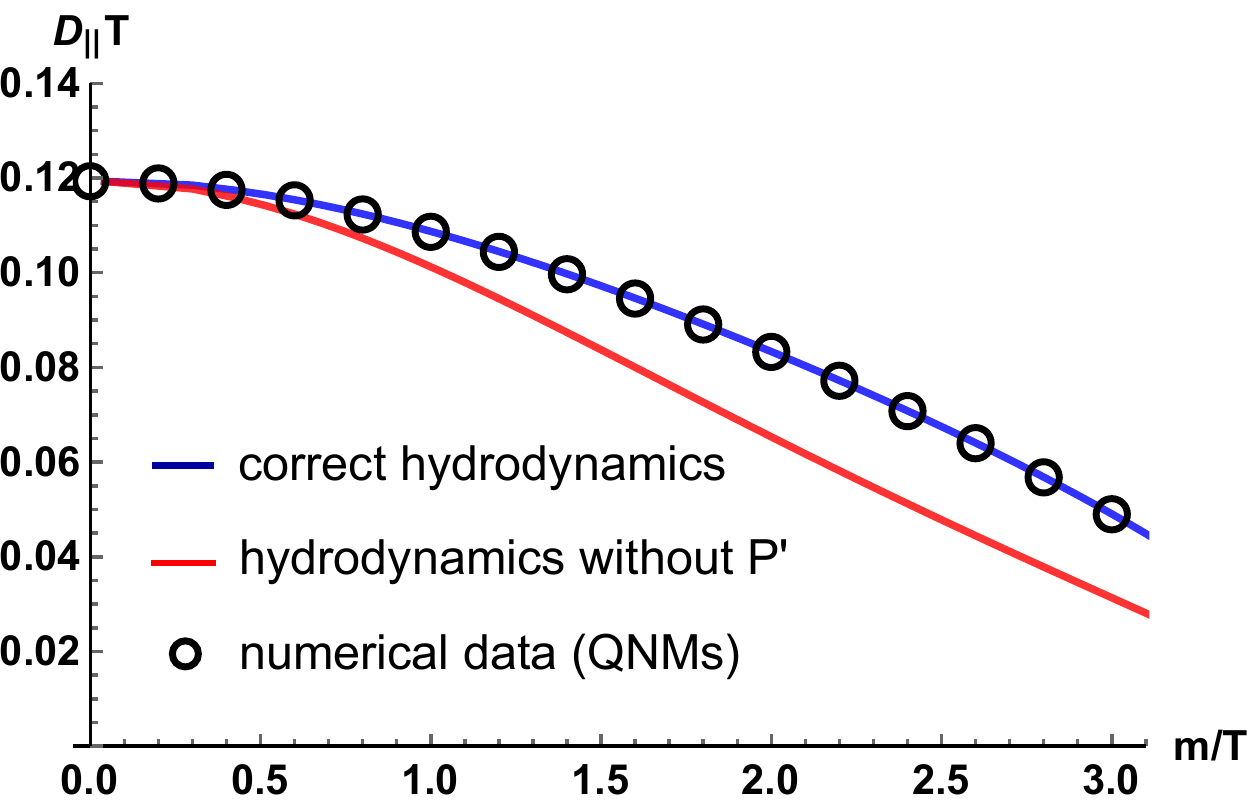}
   \caption{\textbf{Left: }$v_\perp^2$ for $V(X)=X+X^2/2$ model with $\mathcal{P}=0$ alongside the hydrodynamic predictions (solid lines). We have chosen $u_h = 1$ setting $\alpha = 1$. \textbf{Right: }$D_\parallel$ for $V(X)=X+X^2/2$ model with $\mathcal{P}=0$ alongside the hydrodynamic predictions (solid lines). We have chosen $u_h = 1$ setting $\alpha = 1$.}
   \label{fig:unstablevperp}
\end{figure}

We can also compute the quasinormal modes for this system numerically and compare them against the hydrodynamic predictions presented in eq. \eqref{modes_zeroP}, and that of \cite{Delacretaz:2017zxd} without $\cP'$. We see in fig. \ref{fig:unstablevperp} that the transverse speed of sound $v_\perp$ is imaginary due to negative shear modulus $G$; nevertheless the prediction from hydrodynamics matches perfectly.
We again find a discrepancy in $D_\parallel$ similar to \cite{Ammon:2019apj} compared to \cite{Delacretaz:2017zxd}, which is resolved by including $\cP'$ contributions, as in eq. \eqref{modes_zeroP}; see fig. \ref{fig:unstablevperp}.

Despite the simplicity and linear instability of this model, it shares various features of interest with similar holographic models without background strain, such as the one discussed in \cite{Amoretti:2017frz}. Similar models can also be constructed in the frameworks of~\cite{Donos:2012wi,Baggioli:2018bfa,Amoretti:2017frz,Andrade:2017cnc,Donos:2019tmo}. The requirement of thermodynamic stability for isotropic models can be implemented as $\Omega = - \langle  T^{xx} \rangle$ ~\cite{Donos:2013cka}, which according to \eqref{kubos} is precisely $\cP = 0$. Irrespective of the particular model at play, while we might be able to set $\cP = 0$ by judiciously choosing $\alpha$ in the equilibrium state, we will generically be left with a non-zero $\cP'$, which must be taken into account in the dual hydrodynamic theory.\footnote{See also \cite{Donos:2019hpp} for a bulk analysis.}

At this stage, we are not aware of any massive gravity or Q-lattices models which are both thermodynamically and dynamically stable.\footnote{Preliminary results suggest that the model in section II-B of \cite{Amoretti:2017frz} is dynamically unstable as well \cite{private}. This is somewhat expected given the similarities with our $V(X)=X+\lambda X^2$ model.}

\section{Conclusions}

In this paper we illustrated that the theory of viscoelastic hydrodynamics formulated in~\cite{Armas:2019sbe} is the appropriate hydrodynamic description for the (strained) homogeneous holographic  models of~\cite{Alberte:2015isw} with spontaneously broken translations. We showed that the theory faithfully predicts all the transport coefficients and the behaviour of the low-energy quasinormal modes in the holographic setup. Moreover, it resolves the tensions between the previous hydrodynamic framework of \cite{Delacretaz:2017zxd} and the holographic results reported in~\cite{Ammon:2019apj}. 

Moreover, we extended the analysis beyond \cite{Armas:2019sbe} and argued that the effects of the temperature derivative of the strain pressure are present even in unstrained equilibrium configurations. We constructed a thermodynamically stable holographic model, analysed its low-lying QNMs, and found agreement with the expressions in equation \eqref{modes_zeroP}. We have also noted issues (dynamical instabilities) with the physicality of this thermodynamically favoured model (and other similar setups \cite{Armas:2019sbe,Amoretti:2017frz}). 

Generally, we expect that the hydrodynamic formulation of~\cite{Armas:2019sbe}, with the addition of the results and discussions presented in this paper, will continue to work for all homogeneous holographic models with spontaneously broken translations \cite{Amoretti:2017frz, Grozdanov:2018ewh, Donos:2019hpp, Andrade:2017cnc}, due to the same symmetry-breaking pattern.

The analysis in this paper opens up the stage for various interesting future explorations. An immediate goal would be to inspect various holographic models of viscoelasticity in the literature, with zero background strain, and identify the role of non-zero $\cP'$ on the quasinormal spectrum. In particular, the relation between dynamic instability and the absence of strain pressure, which has been presented in this work, is worthy of further investigations. Furthermore, another interesting direction is to better understand the role of strain pressure, and its temperature derivative, in physical systems (see e.g.~\cite{Armas:2020bmo}).

The addition of a small explicit breaking of translations to the hydrodynamic framework of \cite{Armas:2019sbe} could also provide an understanding of the universal phase relaxation relation $\Xi\sim\textsc{M}^2\,/\sigma$ (with $\Xi$ the Goldstone phase relaxation rate; $\textsc{M}$ the mass of the pseudo-Goldstone mode). This relation was proposed in \cite{Amoretti:2018tzw} and was later verified for the models presented in this paper in~\cite{Ammon:2019wci}. It could also provide an explanation for the complex dynamics found in the pseudo-spontaneous limit in \cite{Baggioli:2019abx}. Furthermore, $\Xi$ seems to be tightly connected to the presence of global bulk symmetries, which are not expected to appear in proper inhomogeneous periodic lattice structures. The physical interpretation of these global structures has recently been discussed in \cite{Baggioli:2020nay}, and still represents an important puzzle in the field.

One may also consider the viscoelastic hydrodynamic theory of \cite{Armas:2019sbe} beyond linear response in order to explore the full rheology of the holographic models considered in this work, as initiated in \cite{Baggioli:2019mck}. 

In conclusion, this work marks an important development in understanding the nature of the field theories dual to the widely used holographic models with spontaneously broken translational invariance, and provides another robust bridge between holography, hydrodynamics (in its generalised viscoelastic form) and effective field theory.


\vspace{1em}

\section*{Acknowledgments}
We thank Aristomenis Donos, Blaise Gout\'eraux, Sean Hartnoll, Christiana Pantelidou and Vaios Ziogas for several helpful discussions and comments. S. Gray would like to thank IFT Madrid for hospitality during the initial stages of this work.  S. Grieninger thanks the University of Victoria for hospitality during the initial stages of this work. MA is funded by the Deutsche Forschungsgemeinschaft (DFG, German Research  Foundation) -- 406235073. MB acknowledges the support of the Spanish MINECO’s ``Centro de Excelencia Severo Ochoa'' Programme under grant SEV-2012-0249. The work of S. Gray has been funded by the Deutsche Forschungsgemeinschaft (DFG) under Grant No. 406116891 within the Research Training Group RTG 2522/1. S. Grieninger gratefully acknowledges financial support by the DAAD (German Academic Exchange Service) for a \textit{Jahresstipendium f\"ur Doktorandinnen und Doktoranden} in 2019. AJ is supported by the NSERC Discovery Grant program of Canada.

\bibliographystyle{elsarticle-num}
\let\oldaddcontentsline\addcontentsline
\renewcommand{\addcontentsline}[3]{}
\bibliography{final}
\let\addcontentsline\oldaddcontentsline


\appendix

\section{Holographic Renormalisation}\label{ap1}

In this appendix we give some details regarding the holographic renormalisation underlying the models discussed in the main text. The bulk action \eqref{action} has to be supplemented with appropriate boundary counter terms to have a well-defined variational principle
\begin{equation}
    \label{eq:counter}
    S_{{\text{counter}}}
    \,{=} \int_{~~\mathclap{u=\epsilon}} \df^{d+1} x \sqrt{-\gamma}
    \left( K 
    \,{-}\, \frac{d}{\ell} 
    \,{+}\, m^2 \bar V(\bar{\mathcal{I}}^{IJ}) \right),
\end{equation}
where $\gamma_{\mu\nu} = \lim_{u\to \epsilon}g_{\mu\nu}$ is the induced metric at the boundary, $K$ is the extrinsic curvature, and $\bar{\mathcal{I}}^{IJ} = \gamma^{\mu\nu}\partial_\mu \phi^I \partial_\nu \phi^J$. $\bar V(\bar{\mathcal{I}}^{IJ})$ is an appropriate boundary potential fixed by requiring that the on-shell action of the black brane solution \eqref{backg} to be finite. For instance, in $d=2$, for $V(X) = X^N$ models with $N>3/2$ we have $\bar V(\bar X) = 0$, while for $N<3/2$ we get $\bar V(\bar X) = \bar X/(3-2N)$, where $\bar X = \frac12\mathrm{tr}\,\bar{\mathcal I}$. For $V(X) = X + \lambda X^2$, we instead find $\bar V(\bar X) = \bar X$. 

Due to its novelty, we will in the remainder of this section mainly focus on holographic renormalisation for $V(X) = X + \lambda X^2$.

To implement spontaneous symmetry breaking for models whose boundary behavior goes as $V(X,Z)\sim X^N,Z^M$ with $N<5/2,\,M<5/4$, one needs to apply \textit{alternative quantisation} for the scalars.\footnote{For potentials $V(X,Z)=X^N,Z^M$ with $N>5/2,\,M>5/4$, one instead needs to follow \textit{standard quantisation} in order to have spontaneous symmetry breaking, as shown in \cite{Alberte:2017oqx}.} More precisely, one needs to deform the boundary theory with a term 
\begin{align}
    \label{eq:trace}
    S_{{\text{alt}}}
    &= \int_{~~\mathclap{u=\epsilon}} \df^{d+1} x \sqrt{-\gamma}\,
    \Pi_I \phi^I\,,
\end{align}
where
\begin{align}
    \Pi_I  
    &= \frac{1}{\sqrt{-\gamma}}\frac{\delta(S+S_{\text{counter}})}{\delta \phi^I} =
    \delta_{IJ}
    \left( V'(X) n^a\partial_a \phi^J
    + \nabla_{(\gamma)}^\mu \left(\bar V'(\bar X) \partial_\mu \phi^J \right) \right).
\end{align}
$\nabla^{(\gamma)}_\mu$ is the covariant derivative associated with $\gamma_{\mu\nu}$ and $n_a$ is the outward pointing normal vector at the boundary. The \eqref{eq:trace} term in the action turns $\Phi^I$ at the boundary into the dynamical operator, while the associated source is now given by the boundary value of $\Pi_I$. We are interested in dual hydrodynamic models in the absence of sources for the scalars. Hence, in alternative quantisation we impose the boundary conditions
\begin{equation}
    \lim_{\epsilon \to 0}\frac{1}{\epsilon^{d+1}} \Pi_I = 0\,.
\end{equation}
Finally, for the metric we always impose the standard boundary conditions 
\begin{equation}
    \lim_{\epsilon \to 0} \epsilon^2 \gamma_{\mu\nu} = \eta_{\mu\nu} \,.
\end{equation}
Note that in the {alternative quantisation scheme} the background profile for the scalars, $\phi^I= \alpha x^I$, is no longer an external source providing the explicit breaking of translations. This is the fundamental reason why models like $V(X)=X\,+\,\dots$, using alternative quantisation \cite{Armas:2019sbe}, realize the spontaneous (and not explicit \cite{Andrade:2013gsa}) breaking of translations.

\clearpage
\end{document}